\documentclass[sigconf]{acmart}

\makeatletter
\renewcommand\@formatdoi[1]{\ignorespaces}
\makeatother

\usepackage{xcolor}
\usepackage{graphicx}
\usepackage{wrapfig}

\AtBeginDocument{%
  \providecommand\BibTeX{{%
    \normalfont B\kern-0.5em{\scshape i\kern-0.25em b}\kern-0.8em\TeX}}}

\setcopyright{none}
\copyrightyear{}
\acmYear{}
\acmDOI{} 

\renewcommand\footnotetextcopyrightpermission[1]{}

\acmConference[BioKDD at the KDD Conference]{BioKDD 2020: 19th International Workshop on Data Mining in Bioinformatics - at KDD 2020 The 26th ACM SIGKDD Conference on Knowledge Discovery and Data Mining}{August 24, 2020}{San Diego, CA}
\acmBooktitle{BioKDD 2020: 19th International Workshop on Data Mining in Bioinformatics - at KDD 2020 The 26th ACM SIGKDD Conference on Knowledge Discovery and Data Mining,
  August 24, 2020, San Diego, CA}

\begin{document}

\title[A Deep Learning Pipeline for Patient Diagnosis Prediction Using Electronic Health Records]{A Deep Learning Pipeline for Patient Diagnosis Prediction \\
Using Electronic Health Records}


\author{Leopold Franz}
\email{lfranz@ethz.ch}
\affiliation{%
  \institution{ ETH Z{\"u}rich}
  \streetaddress{Rämistrasse 101}
  \city{Z{\"u}rich}
  \country{Switzerland}
  \postcode{8006}
}

\author{Yash Raj Shrestha}
\email{yshrestha@ethz.ch}
\affiliation{%
  \institution{ ETH Z{\"u}rich}
  \streetaddress{Rämistrasse 101}
  \city{Z{\"u}rich}
  \country{Switzerland}
  \postcode{8006}
}

\author{Bibek Paudel}
\email{bibekp@stanford.edu}
\affiliation{%
  \institution{Stanford University}
  \city{Stanford, CA}
  \country{USA}
}



\begin{abstract}
Augmentation of disease diagnosis and decision-making in health care with machine learning algorithms is gaining much impetus in recent years. In particular, in the current epidemiological situation caused by  COVID-19 pandemic, swift and accurate prediction of disease diagnosis with machine learning algorithms could facilitate identification and care of vulnerable clusters of population, such as those having multi-morbidity conditions. In order to build a useful disease diagnosis prediction system, advancement in both data representation and development of machine learning architectures are imperative. 

First, with respect to data collection and representation, we face severe problems due to multitude of formats and lack of coherency prevalent in Electronic Health Records (EHRs). This causes hindrance in extraction of  valuable information contained in EHRs. Currently, no universal global data standard has been established. As a useful solution, we develop and publish a Python package to transform public health dataset into an easy to access universal format. This data transformation to an international health data format facilitates researchers to easily combine EHR datasets with clinical datasets of diverse formats. 

Second, machine learning algorithms that  predict multiple disease diagnosis categories simultaneously remain underdeveloped. We propose two novel model architectures in this regard. First, DeepObserver, which uses structured numerical data to predict the diagnosis categories and second,  ClinicalBERT\_Multi, that incorporates rich information available in clinical notes via natural language processing methods and also provides interpretable visualizations to medical practitioners. We show that both models can predict multiple diagnoses simultaneously with high accuracy.
\end{abstract}



\maketitle

\section{Introduction}
Application of data-driven decision automation and augmentation in healthcare is gaining increasing importance, especially with proliferation of conditions such as multi-morbidity, where learning algorithms have considerable value to provide.  {\it Multi-morbidity}, which refers to the co-occurrence of multiple chronic conditions has found a place on   the priority agenda for many  healthcare providers and policymakers~\cite{macmahon2018multimorbidity, king2018multimorbidity, navickas2016multimorbidity}. According to the World Health Organization (WHO), multi-morbidity affects up to 25\% of the population in developed countries and this trend is increasing in low- and middle-income countries as well~\cite{mercer2016multimorbidity}.
Despite the increasing concerns about multi-morbidity, professional caregivers under-diagnose up to 71\% of multi-morbid patients. \cite{multimorbidityDiagnosisSensitivity}, a reason being the physicians  frequently missing the diagnosis of diseases outside their field of specialization.
Anecdotal evidence suggest that clinicians over-diagnose mono-morbid patients in only 7\% of the cases, as they are apprehensive about false or too many diagnosis. 

Data-driven methods could help in facilitating both the speed and accuracy with respect to disease diagnosis, and in particular with diagnosis of multiple health conditions. 
Such a benefit is of crucial importance especially with the current epidemiological situation caused by COVID-19, where diagnosis prediction (e.g, those related to multi-morbidity) could help in identifying and attending to vulnerable population clusters who require higher priority in care and safety measures. 
Data-driven applications in health care rely on effective collection and representation of health care data and on the development of machine learning models that can identify robust patterns in the data in order to make diagnosis predictions.  In both these dimensions, we identify limitations in status quo and contribute to improving the situation with respect to multiple diagnosis prediction.  

Data generation has been expedited by digitization of hospitals with about 50 Petabytes of data being generated worldwide every year. 
Currently, at least 97\% of digitally archived hospital  data remains  underutilized, implying enormous potential for improving data-driven decision-making in hospitals and in the health care sector in general \cite{gii_healthcare_data}. 
To create value with health care data, it must be accurately recorded, represented and securely stored considering privacy laws and other socio-technical concerns. 
Carefully archived and represented electronic records can facilitate extraction of meaningful patterns which in turn augment decision-making for clinicians and patients, as well as for pharmaceutical and insurance companies. 

Despite the significant potential of archived health care data, we face a major challenge to scale data-driven digital solutions for health care caused by the fragmented IT landscape. 
For instance, even within the same hospital, different software solutions are implemented in different departments, creating difficulties with respect to knowledge transfer and learning. 
A Swiss regional hospital on average uses up to 40 different software solutions, within the same function of creating and coordinating EHR data~\cite{angerer2017digital}. 
On the one hand, application of multiple software infrastructures for the same task help prevent vast sensitive health data breaches. 
On the other hand, lack of coherent and universal software and data infrastructure exposes data to multiple points of failure, making the system vulnerable to cyber attacks. 
Additionally, the multitude of data architectures creates hindrance in efficient leveraging of EHR for data-driven solutions.  

As an important first step in mitigating this problematic situation, the non-profit organization Health Level 7 International (HL7)~\cite{hl7} aims to overcome challenges caused by the prevalent lack of coherency with respect to data architectures by developing international health care data standards. 
HL7 products are supported by more than 1'600 member hospitals across 50 countries. 
The organisation has developed multiple formats to exchange health care data, among which a popular one is the Fast Healthcare Interoperability Resources (FHIR) format.

A second crucial element of data-driven healthcare systems is to develop algorithms that are able to extract reliable and robust patterns from data and make accurate diagnosis predictions. 
Robust machine learning models predicting mortality, readmission or the discharge diagnoses can be trained from multiple datasets that are combined using the FHIR format. Models that can predict and diagnose all diseases affecting patients are necessary to correctly diagnose multi-morbid patients. 

\textbf{Contributions.} In this paper, we propose a deep learning pipeline towards addressing both problems \textemdash EHR data fragmentation and multi-morbidity detection. 
As a useful contribution in this direction, \emph{we present a Python package to map the Medical Information Mart for Intensive Care (MIMIC-III) dataset to a flat FHIR format.} The MIMIC-III dataset contains EHR data from 46'520 patients and 58'976 ICU admissions\footnote{A hospital admission refers to a complete hospital stay.} and is widely used in studies involving data-driven solutions for EHR. Our Python package makes it possible to use this dataset in an inter-operable format and to combine it with other EHR datasets in different settings.

As a second important contribution, \emph{we evaluate two different deep learning model architectures to help support clinicians in  diagnosing all medical conditions affecting patients simultaneously}. 
The models utilize both structured and unstructured health data to predict admission diagnosis categories. 
The first proposed model, DeepObserver CNN, uses pre-processed numerical observations from the MIMIC-III Chartevents table to predict admission disease diagnoses. Among these numerical observations are vital sign measurements and lab results. 
The second proposed model, ClinicalBERT\_Multi leverages unstructured clinical notes from the MIMIC-III dataset to predict the diagnosis categories. 
ClinicalBERT\_Multi is also trained to diagnose medical conditions after the first 2 or 3 days of admission, allowing patient treatments to be adapted dynamically along an admission according to the diagnoses.

\section{Related Work}

The MIMIC-III dataset has been converted to the FHIR format in \cite{kemp2019baselinemimic} and \cite{mimic_fhir}. Unfortunately \cite{kemp2019baselinemimic} does not make this transformation publicly available and \cite{mimic_fhir} uses Java and PostgreSQL to transform the dataset. This method is complicated to set up and requires a lot of memory allocation. The Python package developed in this paper keeps the data in a flat hierarchy and can be easily integrated into other Python data science pre-processing steps.

Methods such as \cite{deep_patient}, \cite{allam2019neural}, \cite{Rajkomar2018baselinepreprocess}, \cite{choi2018mime}, and \cite{kemp2019baselinemimic} use structured data from EHRs to extract information, represent events, admissions or patients and to predict certain outcomes. Structured data in health care scenarios is often affected by noise, irregularities and inconsistencies. Numeric data is however simple to handle and input into algorithms. Therefore, numerous pre-processing techniques and machine learning algorithms have already been developed, that can easily extract patterns from noisy, irregular and inconsistent data. 

In contrast, handling unstructured data types such as text is complicated. Extracting information from these data types is therefore significantly more challenging. Prior work in the development of machine learning models from multimodal data \textemdash \ often including textual corpora \textemdash \ have shown promising results using multi-task learning~\cite{baumgartner2018aligning}, bolzman machines~\cite{srivastava2012multimodal},  neural encoders~\cite{pezeshkpour2018embedding}, and logical reasoning~\cite{zhang2019iteratively}, among other methods.

Some recent work have also addressed the utilization of multimodal data in healthcare problems, e.g., through improved coding~\cite{xu2019multimodal}, representation learning~\cite{beam2018clinical}, multi-task learning~\cite{wang2019cross,ding2019effectiveness}, and also by augmenting external data sources~\cite{gijsen2020science}.

Textual data contain very rich information that could benefit prediction algorithms. For instance, textual data in the form of clinical notes provide a rich and detailed account of events, admissions and patients  \cite{clinicalbert}, \cite{biobert}, \cite{Rajkomar2018baselinepreprocess}, \cite{kemp2019baselinemimic}. A hospital generates various types of clinical notes during an admission, such as: Radiology Report, Nursing Progress, Physician Report, Echo Report, Discharge Summary and Pharmacy Note. These notes include symptom descriptions, reasons for diagnosis, patient activities and patient histories. Clinicians take a considerable amount of time to read through these notes and interpret a holistic picture of the patients state.

Medical notes are full of abbreviations, jargon and have unusual grammatical structures. Developing models that can represent and learn from the content of clinical notes is challenging. This natural language understanding task is popularly known as representation learning and is already very complex for free-text outside of the medical sector. Multiple approaches have been tested to represent clinical notes such as using a bag-of-words model \cite{bagofwords_zhang2010}, adapted word2vec representations \cite{word2vec_mikolov2017advances}, trained Long Short-term Memory (LSTM) \cite{lstm_gers1999learning} models in \cite{boag2018s} and \cite{Rajkomar2018baselinepreprocess}, hierarchical attention LSTMs in \cite{kemp2019baselinemimic}, Latent Dirichlet Allocation \cite{deep_patient} and multilevel medical embeddings \cite{choi2018mime}. Recent work have also adapted the Bidirectional Encoder Representations from Transformers (BERT) \cite{devlin2018bert} model developed by Google to create context-aware text embeddings. BioBERT \cite{biobert} pre-trains this BERT model with academical biomedical literature and the original BERT training corpus. BioBERT therefore lacks training on actual clinical notes, which have a different structure and more abbreviations. \cite{alsentzer2019publicly} and \cite{clinicalbert} augment the pre-training by adding actual de-identified clinical notes from the MIMIC-III dataset to the training corpus. This addition improves the BERT's capability to create deep clinical note representations. 

Several methods address the diagnose the primary disease of admissions \cite{kemp2019baselinemimic}, \cite{deep_patient}. Some methods also diagnose specific singular diseases, \cite{allam2019neural, norgeot2019assessment}. Meanwhile, \cite{Rajkomar2018baselinepreprocess} and \cite{kemp2019baselinemimic} are developed to additionally predict multiple disease codes simultaneously. More precisely, these models are designed to predict International Classification of Diseases, Ninth Revision, Clinical Modification (ICD-9-CM) codes. In contrast, the models proposed in the current paper simultaneously predict all disease categories affecting the patients using popularly known  Clinical Classifications Software (CCS) categories.

\section{FHIR Transformation}
In this section, we present the  FHIR format and how the open source Python package we developed converts the MIMIC-III dataset into a flat FHIR format.

\subsection{FHIR}
The FHIR format is a new format to exchange EHR data. The FHIR format represents data in containers that are consistent, scalable, and hierarchical. This facilitates the exchange of data between participants in health ecosystems. Moreover, the format does not ensure semantic consistency, which allows the format to be dynamically adjusted and implemented by any health system.

Every recorded event along a patient's admission is considered as a resource in the FHIR format. These resources contain multiple attributes. The ``Medication Dispense'' resource for example contains the trade name, the generic name or the medication ingredients as attributes. Each attribute is defined by a specific data structure and type, creating the hierarchical structure of the resources. These structures and requirements are explicitly documented on the FHIR website \cite{hl7_fhir}. 

Each FHIR resource type belongs to a specific thematic category, which in turn belongs to a certain level depending on how sensitive and widespread the resource type is. The levels define how consistently the resource types are defined and how tightly they are governed. Resources types from the top levels are the most widespread and support the most common health care transactions.

\subsection{MIMIC-III to FHIR Transformation}
To be able to combine the MIMIC-III dataset with other EHR datasets in future applications, we mapped it to the FHIR format. Consequently, all primary MIMIC-III tables were converted to the appropriate FHIR resource type, if a corresponding FHIR resource type existed \footnote{No correspoding FHIR resource type were found for the 'callout', 'transfers' or 'drg\_codes' tables.}, see Table \ref{table:fhir_mapping} for the corresponding mapping. 

\begin{table}[!ht]
	\centering
	\footnotesize
	\begin{tabular}{|c l l|}
 
		\hline
		 & MIMIC-III table & FHIR Resource Type \\ [0.5ex]
		\hline
		1 & patients &  patient\\
		2& admissions & encounter \\
		3& diagnosis\_icd &  encounter\\
		4& icustay & enounter\\
		5&cptevents & claim\\
		6&noteevents & diagnosticReport\\
		7&inputevents\_cv & medicationDispense\\
		8&inputevents\_mv & medicationDispense\\
		9&prescriptions & medicationRequest\\
		10&chartevents & observation\\
		11&datetimeevents & observation\\
		12&labevents & observation\\
		13&caregivers & practitioner\\
		14&procedures\_icd & procedure\\
		15&procedureevents\_mv & procedure\\
		16&microbiology & specimen\\
		17&outputevents & specimen\\
		18&service & serviceRequest\\
		19&callout & \\
		20&transfers & \\
		21&drgcodes & \\
		\hline
	\end{tabular}
	\label{table:fhir_mapping}
	\caption{MIMIC-III Data Mapping to FHIR Resources}
\end{table}

The MIMIC-III data tables correspond to level 3 and 4 FHIR resource types. Instead of saving the FHIR resources as hierarchical containers, the resources are saved as single level containers that can easily be read into Python Pandas DataFrames. 

The FHIR format allows resources to  be saved in multiple file formats such as XML, JSON and Turtle. The proposed Python package saves the FHIR resources as JSON files, as these can be easily read by the Python programming language. Each MIMIC-III table was therefore converted to a collection of FHIR resource objects, saved as a GZIP compressed JSON file.
We provide a visual representation of the admission table entries mapped to the encounter resource type with greater details in the Appendix. 

The python package can be downloaded from its online repository~\footnote{https://github.com/leopold-franz/MIMIC-III\_FHIR\_Transformation}. The next steps for its use are as follows: (i) import the transform function from the mimic\_fhir\_transformation.py, and (ii) call the transform function with these inputs: \textit{input\_path} of the original MIMIC-III table CSV file, \textit{output\_path} of where the collection of FHIR resources should be saved as a JSON file. Note that by adding the '.gz' extension the function can read compressed CSV files and save compressed JSON files. The function then saves the FHIR collection as a file and returns a dataframe with the flat hierarchy FHIR resources as rows. This allows the function call to be directly incorporated into any python pre-processing step. The MIMIC-III FHIR collections can then be combined with FHIR collections from other EHR datasets and used to train robust machine learning models.

\section{Patient Diagnosis Prediction}
This section presents the deep-learning model architectures that we use for disease diagnosis prediction. 

\subsection{DeepObserver}
DeepObserver is a deep learning model inspired by \cite{kemp2019baselinemimic}. In this model, all numerical observations from the MIMIC-III Chartevents table, that contains vital sign measurements and laboratory results, are used for training instead of using only a few variables.  

\subsubsection{Data \& Pre-processing}
\begin{figure}[!ht]
	\centering
	\includegraphics[width=0.45\textwidth]{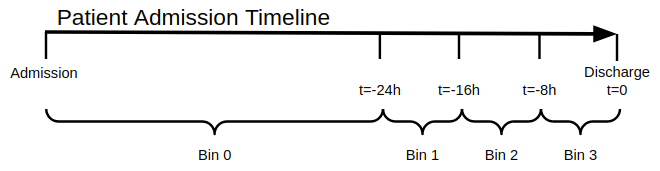}
	\caption{Grouping numerical observations into bins.}
	\label{figure:deepobserver_binning}
\end{figure}

The Chartevents table contains more than 330 million measurements. These measurements belong to different observation types, for example some Chartevent entries denote arterial blood pressure measurements and others denote body temperature data. Depending on the observation type, the measurements might not be numerical values, instead  short strings or DateTime objects. As a first step of pre-processing, all values from non-numerical observation types were filtered out and  only remaining 450 observation types  with numerical values were used. Models in  \cite{kemp2019baselinemimic}, which has similar pre-processing steps, also uses categorical string values converted into numeric embeddings. Next, the time difference between each measurement and the corresponding admission discharge time is computed. Subsequently, all observations are grouped together into 4 time interval  \textit{bins} similar to \cite{kemp2019baselinemimic}. The last three bins correspond to the three consecutive 8h intervals before the discharge time point and all values before the last 24h of an admission are grouped together in the first bin, see Figure \ref{figure:deepobserver_binning}. 

Measurements of the same observation type in the same bin are averaged together. The resulting data of a single admission corresponds to  450 numeric observation types each with 4 time bin values. All values of the same type are then normalized to Z-values.

The input for the DeepObserver model corresponds to an array of size (450,4) filled with normalized numerical observations, representing a patient admission.  The model is trained on 29'590 admissions, as not every MIMIC-III patient admission has numerical entries in the Chartevents table. 

DeepObserver is a supervised model that predicts multiple disease diagnoses. 
Each admission representation is therefore passed to the model with corresponding disease diagnosis labels. There are two locations to record the admission diagnoses within the MIMIC-III dataset. 
First location corresponds to a free-text entry in the Admission table and the second corresponds to the Diagnoses ICD table, where each admission is attributed with ICD-9-CM codes that denote the official diagnosis codes for hospital utilization in the United states. There exists 15'073 ICD-9-CM codes, of which the MIMIC-III dataset contains 6'984. Due to their high number, these ICD-9-CM codes are grouped and mapped to 281 meaningful CCS categories. "The Clinical Classifications Software (CCS) is a tool for clustering patient diagnoses and procedures into a manageable number of clinically meaningful categories." \cite{icd9_ccs_mapping}. There are 284 meaningful CCS diagnosis categories.
Each admission representation is labeled with an array of size 281. Every entry in this array is a Boolean item indicating whether the admission is diagnosed with an diagnosis code of the corresponding CCS category. A positive entry signifies that the patient is diagnosed with the corresponding disease category for the admission. 

\subsubsection{Sampling} A challenge with working with medical data is that very often we face an unbalanced ratio of positive to negative labels. For example some rare CCS categories appear in less than 10 admissions of the MIMIC-III dataset. This means that at least 29'580 other admissions have a negative binary label for this CCS category. If a model uses all of the training data, it will therefore learn to predict a negative label with a high probability. One solution is to under-sample the majority class, which means that negative samples are excluded from the dataset until an equal number positive and negative samples remains, and then split the dataset into train (80\%), validation (10\%) and test (10\%) partitions. Unfortunately, an equal number of positive and negative samples cannot be reached for each label by under-sampling an extremely unbalanced multi-label dataset. Therefore an iterative train-test-split method \cite{multilabelstrat} is used that maintains the original label distribution in the train, validation and test sets. This method first calculates the label distribution in the original dataset and then calculates the label frequency needed in the desired partitions to maintain the label distribution. Next, in an iterative manner the sample with the rarest label in the original dataset is moved to the partition needing the rare label most to reach the required distribution.

\subsubsection{Model Architecture}
The DeepObserver model consists of a four layer neural network with supplementary dropout layers \cite{srivastava2014dropout} as regularization. The first layer reduces the admission representations dimension by learning patterns across the time dimension. Learning these time behaviours for each observation type is extremely important, and therefore different neural network layer architectures were evaluated for the first layer. The first architecture evaluated was a simple Fully Connected Neural Network (FCNN) layer. The second architecture tested was a Convolutional Neural Network (CNN) layer with filters of size (1,4) spanning across the time dimension. Finally, a simple Recurrent Neural Network (RNN) was evaluated, which is considered effective at learning patterns from sequences. Because this sequence is very short, no Long Short Term Memory (LSTM) or GRU (Gated Recurrent Unit) RNN architectures were considered. The second and third layer of the DeepObserver Network are all fully connected neural network layers of size 512. The last layer is a fully connected neural network layer with an output size equal to the number of CCS diagnosis categories to predict and with a sigmoid activation function that transforms the each output into label probabilities. 



\subsubsection{Training}
The DeepObserver model was trained for three epochs with the Adam optimizer \cite{kingma2014adam}, a learning rate of $2e-5$ and a batch size of 32. Increasing the number of epochs did not improve the model. The hardware used for training and evaluating DeepObserver was a machine with one GPU having 11GiB of memory and 2 CPUs nodes with 40GB of RAM. 

\subsection{ClinicalBERT\_Multi}
We develop an adaptation over the ClinicalBERT model \cite{clinicalbert} called ClinicalBERT\_Multi, which predicts disease diagnoses instead of readmission. The ClinicalBERT architecture is able to transform clinical notes into embeddings that can provide interpretable clinical insights. Similar to \cite{clinicalbert}, our adapted model is trained only using discharge summaries and clinical notes from the first 3 days of an admission. A model trained on notes from the first 3 days of admission can predict disease diagnoses early on in the admission, when the patient is still in the hospital. This early model is therefore evaluated on clinical notes from the first 3 and first 2 days of admission. Early predictions are exceedingly valuable in a clinical setting, as they allow patient treatments to be adjusted according to the predicted diagnoses while the patients are still in the hospital. 

\subsubsection{Data Pre-processing}
\label{section:clinicalbert_preprocessing}
The clinical notes are first pre-processed, lower cased and tokenized. During pre-processing some key abbreviations are replaced (e.g., 'dr.' -> 'doctor'), and some superfluous characters are removed (e.g., the new line character '\textbackslash n').

The ClinicalBERT\_Multi model is not trained with all clinical notes that exist per admission. This differs from the DeepObserver model that does use all available numerical observations per admission. Instead, similar to the original ClinicalBERT \cite{clinicalbert}, the clinical notes dataset is split into three subsets are as follows:
\begin{enumerate}
    \item[$D_{disch}$] contains all MIMIC-III discharge summaries. Discharge summaries contain a summary of the admission, a written form of the diagnosis, a list of prescribed medications and even summaries of other clinical notes, for example a radiology report summary. Each admission has only one discharge summary.
    \item[$D_{3days}$] contains all MIMIC-III clinical notes except discharge summaries from the first three days of admission. Clinical Notes other than discharge summaries address a specific issue or service and can be more comprehensive.
    \item[$D_{2days}$] contains all MIMIC-III clinical notes except discharge summaries from the first two days of admission. This subset is only used as an additional evaluation subset of the model trained on $D_{3days}$.
\end{enumerate}

$D_{disch}$ and $D_{3days}$ are then partitioned into 80\% train, 10\% validation and 10\% test sets using the same iterative train-test-split method \cite{multilabelstrat} as DeepObserver. One ClinicalBERT\_Multi model then trains on the train set of $D_{disch}$ and another ClinicalBERT\_Multi model is trained on the train partition of $D_{3days}$.  A model trained to represent notes from the first three days of admission can also represent notes from the first two days of admission. It is therefore not necessary to train an extra model using only notes from the first two days. Consequently all clinical notes in $D_{2days}$, that also exist in the test partition of $D_{3days}$ are saved as a test partition of $D_{2days}$.The $D_{2days}$ subset is then only used as an additional test set of the model trained on $D_{3days}$. In Table \ref{table:subtasks}, information on the different subsets is shown. 

All notes from $D_{3days}$ or $D_{2days}$ belonging to the same admission are concatenated together to form one long text string for each admission. 

The ClinicalBERT architecture \cite{clinicalbert} can however only take in 512 inputs at a time. Therefore, long text sequences are split into multiple smaller text chunks. The text chunk strings are then converted to token sequences using the BERT tokenizer. The BERT tokenizer splits a string into word units and additionally splits rare or long word units into subword units. Furthermore, this tokenizer adds an extra token at the beginning of each sequence, which after being embedded by BERT can be used to classify the whole sequence.

Our models have a different task than the original ClinicalBERT model. The binary readmission labels are therefore transformed to diagnosis array labels, that are identical to the DeepObserver labels.

\subsubsection{Model Architecture}
The ClinicalBERT\_Multi model has the same base architecture as the original ClinicalBERT model, which consists of the standard BERT architecture with an extra fully connected neural network layer on top. 

To understand BERT we must understand how attention and transformers work. Attention assigns weights to every input feature based on the features importance for the task at hand. By visualizing these weights, the learned importance of the input tokens can be shown. This ability makes the model output interpretable which is highy relevant in the clinical setting where clinicians demand models to be interpretative.  Transformer encoders then use this attention to simultaneously process every element of a sequence into a sequence of deep embedding representations. BERT is based on 12 stacked transformer encoders \cite{vaswani2017attention} with 12 attention heads. The number of attention heads in BERT defines how many attention mechanisms are used per transformer, thereby defining how many filters are learned per transformer. 

The last layer after BERT has an output size equal to the number of labels and a sigmoid activation function, which transforms each output into a probability. Consequently, the last layer acts as a classification layer. It is also where the ClinicalBERT\_Multi model differs from the original ClinicalBERT model, as the output size of ClinicalBERT\_Multi's classification layer is not equal to one but to the number of diagnosis categories to predict.

BERT now takes in a text chunk token sequence $\vec{T}$ as input and returns the embeddings of every token in the sequence. As mentioned in the pre-processing step a classification token is inserted at the beginning of every sequence. The last fully connected layer with weights $W$ then uses this embedded classification token  $h_{\text{CLS}}$ output by BERT to compute a diagnosis probability $P$ for each CCS category, see equation \ref{equation:classification layer}. 
\begin{equation}
P(\verb|CCS category 1| = 1 | h_{\text{CLS}}) = \sigma(Wh_{CLS})
\label{equation:classification layer}
\end{equation}
Each probability determines whether the admission will be diagnosed with a disease from the corresponding CCS category. 
\begin{equation}
	ClinicalBERT(\vec{T}) = \begin{bmatrix} P(\verb|CCS category 1| = 1 | h_{\text{CLS}}) \\ P(\verb|CCS category 2| = 1 | h_{\text{CLS}}) \\ ... \\ P(\verb|CCS category 281| = 1 | h_{\text{CLS}}) \end{bmatrix}
	\label{equation:clinicalbert_input_output}
\end{equation}
Equation \ref{equation:clinicalbert_input_output} shows the ClinicalBERT\_Multi input token sequence $\vec{T}$ and text chunk probability array output.a

Multiple text chunks belong to the same admission, therefore the text chunk probabilities of the same admission and same CCS category are combined to a single admission CCS probability using Equation \ref{equation:chunk_probability_combination} from \cite{clinicalbert}, where $h_{admission}$ represents the admission representation. 
\begin{equation}
P(\verb|CCS category x| | h_{admission}) = \frac{P^n_{max} + P^n_{mean}*n/c}{1 + n/c}
\label{equation:chunk_probability_combination}
\end{equation}
This calculated admission level diagnosis category probability improves on the individual disease category probabilities. $P^n_{max}$ is the maximum probability of all text chunks belonging to the same admission and is used to give additional weight to certain text chunks that have a high probability. Analogously $P^n_{mean}$ is the mean probability of all text chunks and is used to attenuate noise from a certain text chunks. $n$ represents the number of text chunks that belong to the corresponding admission. $c$ is a scaling factor that balances the weight given to the maximum probability and the mean probability. $c=2$ was used like \cite{clinicalbert}

Results presented in this paper are based on the admission level diagnosis category probabilities instead of on the text chunk diagnosis probabilities, as the admission level probabilities have shown to "consistently outperform predictions on each subsequence individually by 3-8\%"\cite{clinicalbert}. The resulting admission CCS probabilities can be converted into positive or negative predictions by setting a certain threshold for each CSS category. Any admission CCS probability over the CCS category threshold is converted to a positive prediction otherwise the admission CCS probability is converted to negative prediction. \footnote{Model performance can be evaluated without setting a threshold, which is why such a threshold was not chosen.} 

\subsection{Training}
The pre-trained model from \cite{clinicalbert} was used and fine-tuned to the diagnosis prediction task for one epoch with the Adam optimizer \cite{kingma2014adam} using the standard learning rate of $2e-5$ and a batch size of 8. 
The training and evaluation of the ClinicalBERT\_Multi model was done on a Google Cloud Platform Virtual Machine Instance was setup with 8 CPUs, each having 30GB of memory, 1 NVIDIA Tesla K80 GPU and a 128GB SSD harddisk.

\section{Experiments and Results}
\subsection{Experiments}
The first experiment aims to evaluate the best performing DeepObserver model. Therefore evaluating which architecture for the first layer works best. The different DeepObserver models are DeepObserver FCNN, DeepObserver CNN and DeepObserver RNN, where the second part of the name defines what architecture is used for the first layer.

The second experiment evaluates the performance of different models predicting a specific CCS category: CCS Category 98, which corresponds to Hypertension, also known as high blood pressure. This particular category was chosen as it is the most frequent CCS category in the MIMIC-III dataset, appearing 21'139 times. Results of this experiment are labelled with 'CCS Cat:98' in the task column of Table \ref{table:results}. The models evaluated in for this subtask are DeepObserver CNN, ClinicalBERT\_Multi and additionally ClinicalBERT\_Binary. ClinicalBERT\_Binary is a replica of the original ClinicalBERT model only the task is changed to predicting CCS category 98, which means that each binary readmission label is changed to a binary label defining whether the admission is diagnosed with the CCS category 98.

The goal of the third experiment is to compare the performance of the ClinicalBERT\_Binary and ClinicalBERT\_Multi when evaluated on the different data subsets $D_{disch}$, $D_{3days}$, $D_{2days}$. This experiment therefore evaluates how well the models perform when predicting the diagnosis categories at different admission time points, i.e. after 2 days of admission, after 3 days or after discharge.

The fourth and main experiment compares the predictive performance of all models predicting all diagnosis categories simultaneously. Therefore comparing ClinicalBERT\_Multi using Discharge Summaries, ClinicalBERT\_Multi using Clinical notes from the first 3 days,  ClinicalBERT\_Multi using Clinical notes from the first 2 days, DeepObserver FCNN, DeepObserver CNN and DeepObserver RNN and the SHiP model on the multi-label prediction task.

\subsection{Results}
\label{section:results}
In Table \ref{table:results}, we present the evaluation of all models using AU-ROC, AU-PR, and Recall at $precision=80\%$ (Recall@Prec80). Note that the AU-ROC score is only representative of the predictive performance of models evaluated on balanced datasets such as the ClinicalBERT\_binary and the original ClinicalBERT. Therefore, to be able to compare all models, the AU-PR scores are compared, as well as the recall@prec80 scores. All multi-label classification models are marked with $ ^{b} $ to signify that the micro averaged scores are shown. 

In the case of disease diagnosis support systems for multi-morbid patients, it is important to develop models with low miss rates. High recall scores are therefore required to make sure a few diseases are missed. Low precision scores are not very problematic, as high numbers of false alarms will be disregarded by doctors with a high precision rate.

The AU-PR score of a random classifier is equal to the ratio of positive samples in the evaluation dataset. The ratio of positive CCS Cat.:98 samples in the test partition of the Diagnoses task is 0.071. Therefore, if ClinicalBERT\_Multi and DeepObserver reach AU-PR scores above 0.071 in the CCS Cat.:98 , they outperform a random classifier. Meanwhile, the ClinicalBERT\_Binary model is tested with an equal amount of positive and negative samples, due to the original ClinicalBERT model under-sampling negative samples. Consequently, the AU-PR score of ClinicalBERT\_Binary model should be above 0.5 to outperform the random classifier. Note that the AU-PR scores of all models assessed on the CCS Cat.:98 task can still be compared. The ratio of positive labels in the Diagnoses task is 0.043, which means any Micro-AU-PR score in the Diagnoses task above this value outperforms a random classifier.

\begin{table}[!ht]
	\centering
	\begin{tabular}{l c c c c}
		\hline
		Model & Dataset & \# of Pat. & \# of Adm. & Avg. chunks \# \\ [0.5ex]
		\hline
		ClinicalBERT & $D_{disch}$ & 24'742 & 29'974 & 4.89\\
		\_Binary& $D_{3days}$ & 25'103 & 30'476 & 8.54\\
		& $D_{2days}$ &  &  & 6.28\\
		ClinicalBERT & $D_{disch}$ & 27'238 & 33'684 & 4.64\\
		\_Multi& $D_{3days}$ & 27'564 & 34'152 & 7.31\\
		& $D_{2days}$ &  &  & 5.51\\
		\hline
	\end{tabular}
	\caption{\footnotesize{Details of ClinicalBERT datasets. `\# of Pat.' and `\# of Adm.' show how many patients and admissions are in the train partition of the data subsets. `Avg. chunks \#' indicates the average number of text chunks per admission in the evaluation partition.}}
	\label{table:subtasks}
\end{table}

\begin{table}[!ht]
    \footnotesize
    \centering
	\begin{tabular}{l l l c c c} 
		\toprule
		\textbf{Model} & \textbf{Task} & \textbf{$t_{pred}$} & \textbf{AU-ROC}  & \textbf{AU-PR} & \textbf{Recall@} \\
		            &       &           &          &    & \textbf{Prec80} \\ [0.5ex] 
		\midrule
		\textbf{SHiP $ ^{a} $} & \textit{Diagnoses} &\textit{Discharge} & \textit{0.897} & \textit{0.352} &\textit{ -} \\
		\midrule
		\textbf{DeepObserver} & Diagnoses & Discharge & 0.897$ ^{b} $ & 0.367$ ^{b} $ & 0.039$ ^{b} $\\
		\hspace{8mm} \textbf{FCNN}         &           &           &               &               &              \\
		\midrule
		\textbf{DeepObserver} & Diagnoses & Discharge & 0.866$ ^{b} $ & 0.231$ ^{b} $ & 0.000$ ^{b} $\\
		\hspace{8mm} \textbf{RNN}         &           &           &               &               &              \\
		\midrule
		\textbf{DeepObserver} & Diagnoses & Discharge & 0.90$ ^{b} $ & 0.372$ ^{b} $ & 0.039$ ^{b} $\\
		\cmidrule(lr{1em}){2-6}
		\hspace{8mm} \textbf{FCNN} & CCS Cat: 98 & Discharge & 0.699 & 0.585 & 0.001\\
		\midrule
		\textbf{ClinicalBERT\_} & CCS Cat.: 98 & Discharge & \textbf{0.873} & \textbf{0.862} & \textbf{0.803}\\
		\cmidrule(lr{1em}){3-6}
		\hspace{8mm} \textbf{Binary} &  & 3 Days & 0.751 & 0.726 & 0.219\\
		\cmidrule(lr{1em}){3-6}
		&  & 2 Days & 0.745 & 0.720 & 0.239\\
		\midrule
		\textbf{ClinicalBERT\_} & Diagnoses & Discharge & 0.919$^{b}$ & 0.408$^{b}$ & 0.08$1^{b}$\\
		\cmidrule(lr{1em}){3-6}
		\hspace{8mm} \textbf{Multi} &  & 3 Days & \textbf{0.921}$ ^{b} $ & \textbf{0.426}$ ^{b} $ & \textbf{0.117}$ ^{b} $\\
		\cmidrule(lr{1em}){3-6}
		&  & 2 Days & 0.708$ ^{b} $ & 0.202$ ^{b} $ & 0.061$ ^{b} $\\
		\midrule
		\textbf{ClinicalBERT\_} & CCS Cat.: 98 & Discharge & 0.739 & 0.626 & 0.032\\
		\cmidrule(lr{1em}){3-6}
		\hspace{8mm} \textbf{Multi} &  & 3 Days & 0.723 & 0.613 & 0.082\\
		\cmidrule(lr{1em}){3-6}
		&  & 2 Days & 0.666 & 0.558 & 0.003\\[1ex] 
		\bottomrule
	\end{tabular}
	\caption{\footnotesize{Predictive qualities of deep learning for disease diagnoses. The prediction time point is represented by $t_{pred}$. The \textbf{Bold values} are the values of the best performing model per task. $ ^{a} $ The Sequential, Hierarchical, and Pretrained Model results are retrieved from \cite{kemp2019baselinemimic} paper, which diagnoses multiple ICD-9 codes. $^{b}$ indicates that the values are micro averaged evaluation values. $ ^{c} $ The ClinicalBERT results are retrieved from the \cite{clinicalbert} paper. \textit{Results in italics are retrieved from the original papers}. Any non italicized values are results from models proposed in this paper.}}
	\label{table:results}
\end{table}

Next, we summarize the results of the predictive performance of models and the results of the experiments. 

All models proposed in this paper perform better than a random guess, as they all have an AU-PR score well above the AU-PR score of a random classifier. The smallest performance improvement compared to a random classifier for models evaluated on an imbalanced dataset is the ClinicalBERT\_Multi model. It predicts the disease diagnosis codes after \textbf{2} days of admission with a performance improvement of nearly 370\% (0.202/0.043 -1). The ClinicalBERT\_Binary model is the only model trained for this paper, which trains with an equal ratio of positive and negative samples. It shows significant improvement from the random classifier  of 50\%. 

The results of the first experiment show that the best performing DeepObserver model is the DeepObserver CNN model with a Micro-AU-PR score of 0.372. This indicates that the CNN filters are best at capturing time dependent patterns in these circumstances. The lower performance of the FCNN model is probably due to the higher number of trainable weights (500\% more weights). The lower performance of the DeepObserver RNN model can be explained by the low number of time steps along the time dimension the RNN can learn patterns from.

For the second experiment, we compare the models performances of models predicting the CCS Cat.:98 task. The best performing model is ClinicalBERT\_Binary predicting the diagnosis of the 98th CCS category at the Discharge time point. It achieves an AU-PR score of 0.862 and an exceptionally high recall@prec80 of 0.803. The ClinicalBERT\_Multi model trained to predict all diagnosis categories at Discharge is evaluated when predicting CCS Cat.:98 and achieves an AU-PR score of 0.626. This shows that fine-tuning the ClinicalBERT architecture on a balanced single diagnosis code dataset compared to on all diagnosis categories achieves a superior performance. 

Results from the third experiment show that the prediction time point, which directly influences the amount of input data, has an effect on the predictive performance of models. The AU-PR scores of ClinicalBERT, ClinicalBERT\_Binary and ClinicalBERT\_Multi models improve when evaluating clinical notes of the first 3 days compared to evaluating clinical notes from the first 2 days. The difference in input data amount can be found in the 'Avg. chunks \#' column of Table \ref{table:subtasks}. All previously mentioned models show a higher AU-PR score when making predictions after 3 days of admission compared to when making predictions after 2 days of admission. Furthermore, ClinicalBERT\_Binary and ClinicalBERT\_Multi evaluated on task 'CCS Cat.: 98' have a better AU-PR score when using the discharge diagnoses than when using all clinical notes from the first 3 days of admission. Interestingly, the ClinicalBERT\_Multi sub-model predicting diagnoses after 3 days (AU-PR = 0.426) slightly outperforms ClinicalBERT\_Multi trained and evaluated with discharge summaries(AU-PR = 0.408). This could be due to the higher amount of input data when making predictions using clinical notes of the first 3 days of admission compared to making predictions using discharge summaries. The difference in input data is seen in column 'Avg. chunks \#' of table \ref{table:subtasks}.

The results of the fourth experiment show that the model with the highest predictive performance for the Diagnoses task is the ClinicalBERT\_Multi model predicting disease diagnosis categories using notes of the first 3 days of admission. It has an AU-PR score of 0.426. It slightly outperforms the DeepObserver CNN model making predictions after discharge (AU-PR = 0.372). This performance difference can be explained by the richer information in clinical notes when compared to the information in the numerical observations. Note that the high accuracy of the DeepObserver CNN model remains a significant achievement, as the data from the DeepObserver model is not human generated and does not depend on quality of clinical notes written by caregivers. Furthermore, comparing the results of the ClinicalBERT\_Multi model predicting CCS categories after 3 days with the results of the external SHiP model (AU-PR = 0.352) shows that ClinicalBERT\_Multi has a higher AU-PR score. The SHiP model differs from the CLinicalBERT\_Multi task by the following two points: (1) it uses all notes from an admission to make diagnosis predictions, which should help the SHiP model to achieve a higher predictive score, (2) the SHiP model predicts ICD-9-CM codes, which are harder to predict as there are more of them, instead of CCS categories.

As a summary, the DeepObserver CNN, which uses numerical observations, easily outperforms a random classifier when predicting multiple diagnosis categories simultaneously. Moreover, ClinicalBERT\_Multi, which uses clinical notes, has an even higher predictive performance than DeepObserver CNN when predicting multiple diagnosis categories and is can even provide good predictions when predicting the diagnoses early on in the admission. 

\subsection{ClinicalBERT Interpretation}
By visualizing the self-attention heads from BERT, the diagnoses predictions of ClinicalBERT\_Binary and ClinicalBERT\_Multi can be interpreted. The attention function takes as input a key, a query and a value vector. Using these vectors it computes a distribution over all keys for each query and then multiplies these distribution weights with the values. In the case of self-attention the queries, keys and values are each constructed by multiplying the input tokens with a set of learned weights. The attention weight distribution for a query vector $q$ of length $ d $, and a set of keys $K$ each also of the length $ d $ is therefore:

\begin{equation}
\text{Attention}(Q, K, V) = \text{softmax}\left( \dfrac{QK^\top}{\sqrt{d}}\right)V
\end{equation}
The self-attention heads therefore act as filters over the input token sequence.

\begin{figure}[!ht]
	\centering
	\includegraphics[width=0.45\textwidth,keepaspectratio]{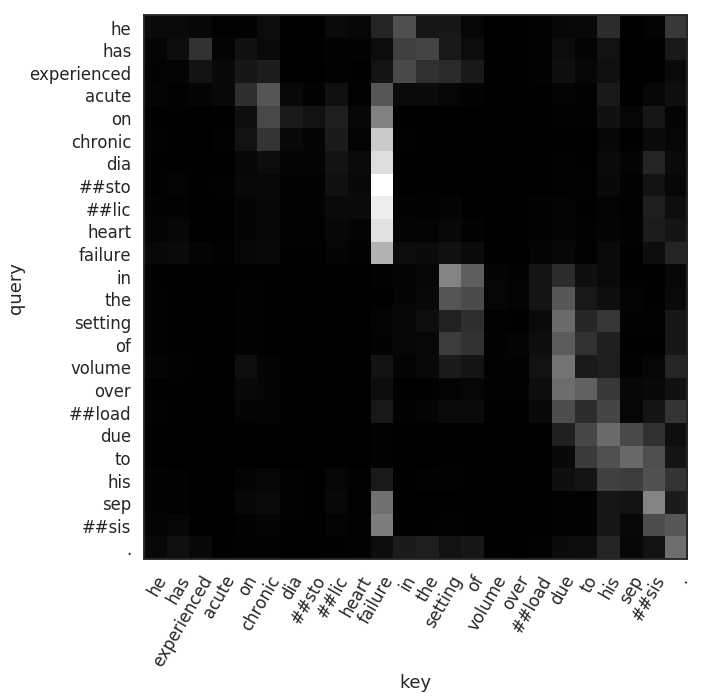}
	\caption{Self-attention weights aligned to key and query tokens from  the ClinicalBERT\_Binary model trained on discharge summaries.}
	\label{figure:attention_short_input}
\end{figure}

An attention weight with a high value signifies that the interaction between the query and key token is predictive about the task at hand. In Figure \ref{figure:attention_short_input}, the weights of a single self-attention head are visualized. As you can see a the failure token is given a high weight.

Being able to interpret the results from ClinicalBERT\_Multi sub-models is important to gain an understanding of the model predictions and better informing medical practitioner's decision-making. 

\section{Conclusion, Limitations, and Future Work}
Our paper contributes a deep-learning pipeline for patient diagnosis prediction using multimodal data sources. We address this problem in two aspects: data representation scheme as well as neural-network model architectures. This paper identifies the problems arising due to multitudes of data representation and describes a Python package to map the Medical Information Mart for Intensive Care (MIMIC-III) dataset which contains EHR data from 46'520 patients and 58'976 ICU admissions to a flat FHIR format as a useful solution. In terms of models introduced models in this paper, i.e., ClinicalBERT\_Multi and DeepObserver architectures come with various advantages. ClinicalBERT\_Multi model which outputs early prediction after 3 days of admission has the highest predictive performance among all proposed models predicting multiple disease diagnosis codes simultaneously. This suggests the effectiveness of  ClinicalBERT\_Multi  in extracting meaningful information from valuable clinical notes. Additionally,  it shows that ClinicalBERT\_Multi is able to make these predictions at significantly early time point. Moreover, this model enables interpretation for medical practitioners by visualisation of the model's self-attention layers. DeepObserver CNN also has a high AU-PR score and does not depend on human written data and  quality. The extension of the diagnosis probabilities of both models in the form of a suitable combination can be explored in future research. The DeepObserver model takes into consideration all numerical observations from the Chartevents table in the MIMIC-III dataset. Future research should aim at augmenting DeepObserver model to learn and use embeddings from all other observation data types in the Chartevents table. 

Despite these contributions, our models also has multiple limitations that provide useful opportunities for future work. 

First, a major limitation of binning observations into the four time bins is the loss of information on short-term irregularities. For instance, conditions such as heart arrhythmia can only be identified when analyzing high resolution heart rate data. Averaging all values within 8h bins removes any such high resolution patterns. Furthermore, being a black-box model in nature, the DeepObserver model is difficult to interpret by medicine practitioners. Future research should aim at producing more interpretable version of DeepObserver. 

Second, ClinicalBERT\_Multi use clinical notes to predict CCS categories, thus inheriting all limitations from the ClinicalBERT and BERT models. One such limitation is the limited input sequence length. ClinicalBERT can only take token sequences with a maximum length of 512 tokens as input and long-term dependencies beyond the 512 tokens are therefore lost. A potential solution that could be considered in future research is to  analyze each clinical note individually instead of concatenating them all together, which could ensure that all text relations in clinical notes shorter than a fixed token size could be correctly captured. 

Third, we directly borrow the intuition based Equation \ref{equation:chunk_probability_combination} proposed by the ClinicalBERT authors to combine the diagnoses probabilities of multiple text chunks. In future, a neural network that is trained to combine these disease diagnosis probabilities could prove useful in performance improvement of the overall disease diagnoses.

An additional limitation with ClinicalBERT\_Multi is that the self-attention heads are not only trained to find important tokens for a single diagnosis category prediction, but learn to identify all relevant tokens for every CCS diagnosis category prediction at the same time. This limits the capacity to interpret predictions. 

Furthermore, both models inherit the usual limitations of retrospective studies. Additionally, the models are only trained on data from a single hospital. As a future step, cross validation of models with data generated from multiple hospitals could be used.  
Overall, we expect a) the proposed FHIR transformation Python package will help researchers in facilitating development of robust machine learning models on EHR datasets, and b) the evaluated deep learning models can contribute towards improved healthcare practice by supporting health care professionals in diagnosing multiple conditions accurately.

\newpage

\bibliographystyle{ACM-Reference-Format}
\bibliography{09_bibliography}

\newpage
\appendix
\section{Appendix}
\subsection{Admission FHIR Mapping}
\includegraphics[scale=0.75, angle=90]{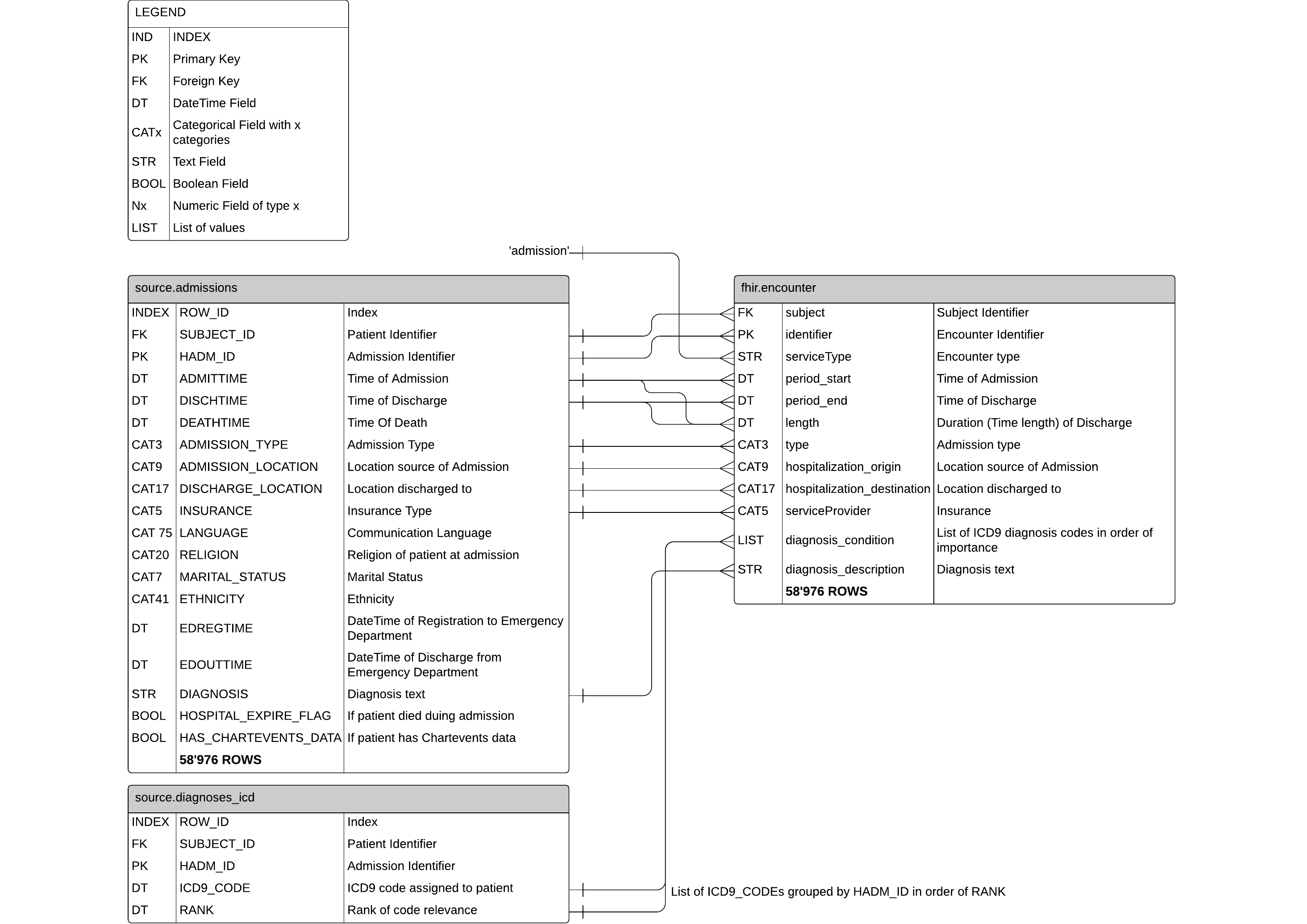}
\end{document}